\documentclass[lettersize,journal]{IEEEtran}
\usepackage{amsmath,amsfonts}

\usepackage{caption}
\usepackage{textcomp}
\usepackage{stfloats}
\usepackage{url}
\usepackage{verbatim}
\usepackage{graphicx}
\usepackage{cite}
\usepackage{multirow,enumitem,adjustbox}
\usepackage{booktabs,threeparttable,makecell}
\usepackage{xcolor}
\usepackage{subcaption}

\usepackage{colortbl}
\usepackage[colorlinks=true,citecolor=green]{hyperref}
\usepackage{cleveref}
\usepackage{rotating}

\usepackage[linesnumbered,ruled,vlined]{algorithm2e}

\SetCommentSty{mycommfont}

\definecolor{lightgray}{RGB}{211, 211, 211}

\begin{document}

\title{Attention-based Encoder-Decoder End-to-End Neural Diarization with Embedding Enhancer}

\author{
Zhengyang Chen, \IEEEmembership{Student Member, IEEE,}
Bing Han, \IEEEmembership{Student Member, IEEE,}
Shuai Wang, \IEEEmembership{Member, IEEE,}
and Yanmin Qian, \IEEEmembership{Senior Member, IEEE,}
}

\markboth{Journal of \LaTeX\ Class Files,~Vol.~14, No.~8, August~2021}%
{Shell \MakeLowercase{\textit{et al.}}: A Sample Article Using IEEEtran.cls for IEEE Journals}


\maketitle

\begin{abstract}
Deep neural network-based systems have significantly improved the performance of speaker diarization tasks. However, end-to-end neural diarization (EEND) systems often struggle to generalize to scenarios with an unseen number of speakers, while target speaker voice activity detection (TS-VAD) systems tend to be overly complex. In this paper, we propose a simple attention-based encoder-decoder network for end-to-end neural diarization (AED-EEND).
In our training process, we introduce a teacher-forcing strategy to address the speaker permutation problem, leading to faster model convergence. For evaluation, we propose an iterative decoding method that outputs diarization results for each speaker sequentially. Additionally, we propose an Enhancer module to enhance the frame-level speaker embeddings, enabling the model to handle scenarios with an unseen number of speakers. We also explore replacing the transformer encoder with a Conformer architecture, which better models local information.
Furthermore, we discovered that commonly used simulation datasets for speaker diarization have a much higher overlap ratio compared to real data. We found that using simulated training data that is more consistent with real data can achieve an improvement in consistency. 
Extensive experimental validation demonstrates the effectiveness of our proposed methodologies. Our best system achieved a new state-of-the-art diarization error rate (DER) performance on all the CALLHOME ($\mathbf{10.08\%}$), DIHARD II ($\mathbf{24.64\%}$), and AMI ($\mathbf{13.00\%}$) evaluation benchmarks, when no oracle voice activity detection (VAD) is used. Beyond speaker diarization, our AED-EEND system also shows remarkable competitiveness as a speech type detection model.
\end{abstract}


\section{Introduction}
\IEEEPARstart{S}{peaker} diarization is a challenging task in speech processing, aiming to determine ``who spoke when'' in scenarios with multiple speakers. It serves as a fundamental pre-processing step in various speech-related tasks. For instance, it enables the detection of distinct speaking segments for each individual present in a recording, allowing subsequent speaker recognition models to identify the absolute speaker identity \cite{jousse2009automatic}. In meeting scenarios, obtaining the speaking segments for each participant is essential to leverage automatic speech recognition (ASR) systems for generating transcripts for individual speakers \cite{kanda2019simultaneous}. Moreover, speaker diarization and speech separation tasks share similarities, leading researchers to explore methods that leverage one task to improve the other \cite{raj2021integration,von2019all,maiti2023eend}.

Conventional speaker diarization systems typically involve multiple stages \cite{shum2013unsupervised,sell2014speaker}. Firstly, a voice activity detection (VAD) system is used to filter the non-speech region and the left part is segmented using a specified window length and hop length. Then, a pre-trained speaker embedding extractor is utilized to extract speaker embeddings for each segment. Subsequently, a scoring backend, such as cosine scoring or PLDA \cite{rajan2014single,villalba2013handling}, is applied to compute similarity scores between pairs of segments. Following this, a clustering algorithm \cite{sell2018diarization,wang2018speaker,lin2019lstm} is employed to assign a unique speaker label to each segment. Optionally, compensation algorithms like Variational-Bayesian refinement \cite{sell2015diarization,sell2018diarization,diez2018speaker,diez2019bayesian} may be employed to refine the clustering results. However, due to the limitations of clustering algorithms, each segment can only be assigned to a single class. This limitation hinders traditional methods from effectively handling scenarios with speaker overlap.

To address the issue of traditional methods being unable to handle speaker overlap, researchers have proposed end-to-end neural diarization (EEND) methods. In \cite{fujita2019end_lstm, fujita2019end_sa}, the authors treated speaker diarization as a frame-wise multi-class classification problem and employed the permutation invariant (PIT) loss \cite{yu2017permutation} to optimize the entire system in an end-to-end manner. However, the number of classes in \cite{fujita2019end_lstm, fujita2019end_sa} is fixed and determined by the output head dimension, limiting their ability to handle scenarios with a flexible number of speakers.
To overcome this limitation, Fujita et al. \cite{fujita2020neural} and Takashima et al. \cite{takashima2021end} proposed a chain-rule paradigm, enabling the sequential output of diarization results for each speaker. This approach allows for flexibility in handling scenarios with varying numbers of speakers. Additionally, Horiguchi et al. \cite{horiguchi2020end} introduced the EEND-EDA system, which utilizes an LSTM encoder-decoder network to model attractors for each speaker. Furthermore, researchers also proposed two-stage hybrid systems \cite{kinoshita21_interspeech, horiguchi2022online} to address the challenge of handling a flexible number of speakers. These systems first output diarization results for short segments with a limited number of speakers using EEND, and then employ a clustering algorithm to solve the inter-segment speaker permutation problem.

Recent research has highlighted the potential benefits of incorporating speaker-specific prior information to enhance system performance and enable the output of speaker-related results. For instance, in speech separation systems \cite{delcroix2018single,xu2020spex,vzmolikova2019speakerbeam}, automatic speech recognition tasks \cite{delcroix2018single}, and active speaker detection tasks \cite{jiang2023target}, researchers have successfully integrated the target speaker's speech or embedding to obtain the desired outputs, such as the target speaker's separated speech, transcript, or on-screen person speaking frames.
Similarly, in the context of speaker diarization, researchers have explored the integration of target speaker prior information, referred to as target speaker voice activity detection (TS-VAD) \cite{medennikov2020target,medennikov2020stc}. To utilize the TS-VAD system for generating results for all speakers, an additional diarization system is often employed to identify the single-speaker speaking segments for each individual. Subsequently, a pre-trained speaker embedding extractor is used to obtain speaker embeddings. Despite the complexity of TS-VAD systems, their excellent performance in competitions \cite{medennikov2020stc,wang2022dku} has motivated researchers to investigate various TS-VAD approaches \cite{cheng2022multi,wang2022target,cheng2022target,jiang2023target}.

As mentioned previously, while the EEND-EDA approach can handle scenarios with a variable number of speakers, the authors in \cite{horiguchi2022encoder} acknowledge that the output speaker number of EEND-EDA is empirically constrained by the maximum number of speakers observed during pre-training. Additionally, the TS-VAD system, with its numerous sub-systems, introduces excessive complexity to the overall system. 
In our previous work \cite{chen2023attention}, we presented a simple attention-based encoder-decoder end-to-end\footnote{In the context of our method, end-to-end refers to the capability of optimizing each module in an end-to-end manner.} neural diarization system (AED-EEND). In this approach, we replace the LSTM encoder-decoder architecture in EEND-EDA with a transformer decoder and achieve better performance. Furthermore, we introduce a teacher-forcing training strategy that leverages speaker-specific prior information. This strategy effectively mitigates the speaker permutation problem and facilitates faster convergence of the system. Additionally, we propose a heuristic decoding method to iteratively obtain diarization results for each speaker.

Due to space limitation, our analysis of the AED-EEND system in our previous article was not sufficient. In this paper, we will give a more comprehensive analysis of it and evaluate the system on more diarization benchmarks to verify its effectiveness. In our previous work \cite{chen2023attention}, we noticed that our AED-EEND system suffers the same problem as EEND-EDA, which has poor performance when the speaker number of the evaluation set is different from the pre-training simulation set. In this paper, we proposed to add a new \textit{Enhancer} module to our AED-EEND system, which can help the model generalize to the unseen number-of-speaker scenario. 
During evaluation on datasets with significantly long durations, we observed that the decoding method with clustering operations, as described in \cite{chen2023attention}, exhibited slow execution. In this paper, we introduce improvements to the decoding method to overcome this challenge. 
Additionally, we discovered a discrepancy between the commonly used simulation dataset in \cite{fujita2019end_lstm,fujita2019end_sa,fujita2020neural,horiguchi2020end} and real data, particularly in terms of the overlap ratio. 
To address this mismatch, we introduce a more realistic data simulation approach following the guidelines outlined in \cite{landini22_interspeech,landini2023multi}.
In contrast to previous experiments \cite{landini22_interspeech,landini2023multi} that focused on limited scenarios and employed relatively weaker systems, Notably, our proposed approach achieves state-of-the-art performance across multiple diarization evaluation benchmarks.
Furthermore, we explore the replacement of the transformer encoder in AED-EEND with the Conformer encoder, leading to further enhancements. 
Although AED-EEND is a diarization system, we discovered that we can also utilize it as a standalone speech type detection model to identify non-speech, single-speaker speech, and overlapping speech regions within the audio. The main contributions of this paper can be summarized as follows:
\begin{enumerate}
    \item We extend our previous conference paper \cite{chen2023attention} by providing an expanded and in-depth analysis, as well as refining the previous flawed decoding method with clustering operations.
    \item We propose a novel Enhancer module to help the frame-level speaker embedding incorporate useful information from the speaker attractor. The experiments show that the Enhancer module can help the system generalize to the unseen number-of-speaker scenario.
    \item We explore the impact of simulation data configuration on our system performance and validate it on multiple datasets.
    \item We compare the system performance when leveraging the transformer encoder or Conformer encoder, and validate it on multiple datasets.
    \item We conduct a thorough investigation of the systems on the relevant datasets. Compared with other diarization systems, we achieve the new state-of-the-art performance on all the CALLHOME, AMI, and DIHARD II evaluation benchmarks when no oracle voice activity information is used.
    \item We also evaluate the proposed AED-EEND system as a standalone speech type detection model. Our evaluation reveals that the proposed AED-EEND system exhibits notable competitiveness in detecting non-speech, single-speaker speech, and overlapping speech regions.

\end{enumerate}

\section{Novel Attention-based Encoder-Decoder End-to-End Neural Diarzation}
In this section, we provide a detailed introduction to our method. Firstly, we describe our designed attention-based encoder-decoder end-to-end neural diarization (AED-EEND) system, highlighting its key components and architecture. Then, we propose the Enhancer module to improve the frame-level speaker embedding representation. And next, we show how we train our system with the teacher-forcing strategy. Finally, we will describe our proposed iterative decoding strategy to output the diarization results for each speaker sequentially.

\begin{figure}[ht!]
  \centering
  \includegraphics[width=0.46\textwidth]{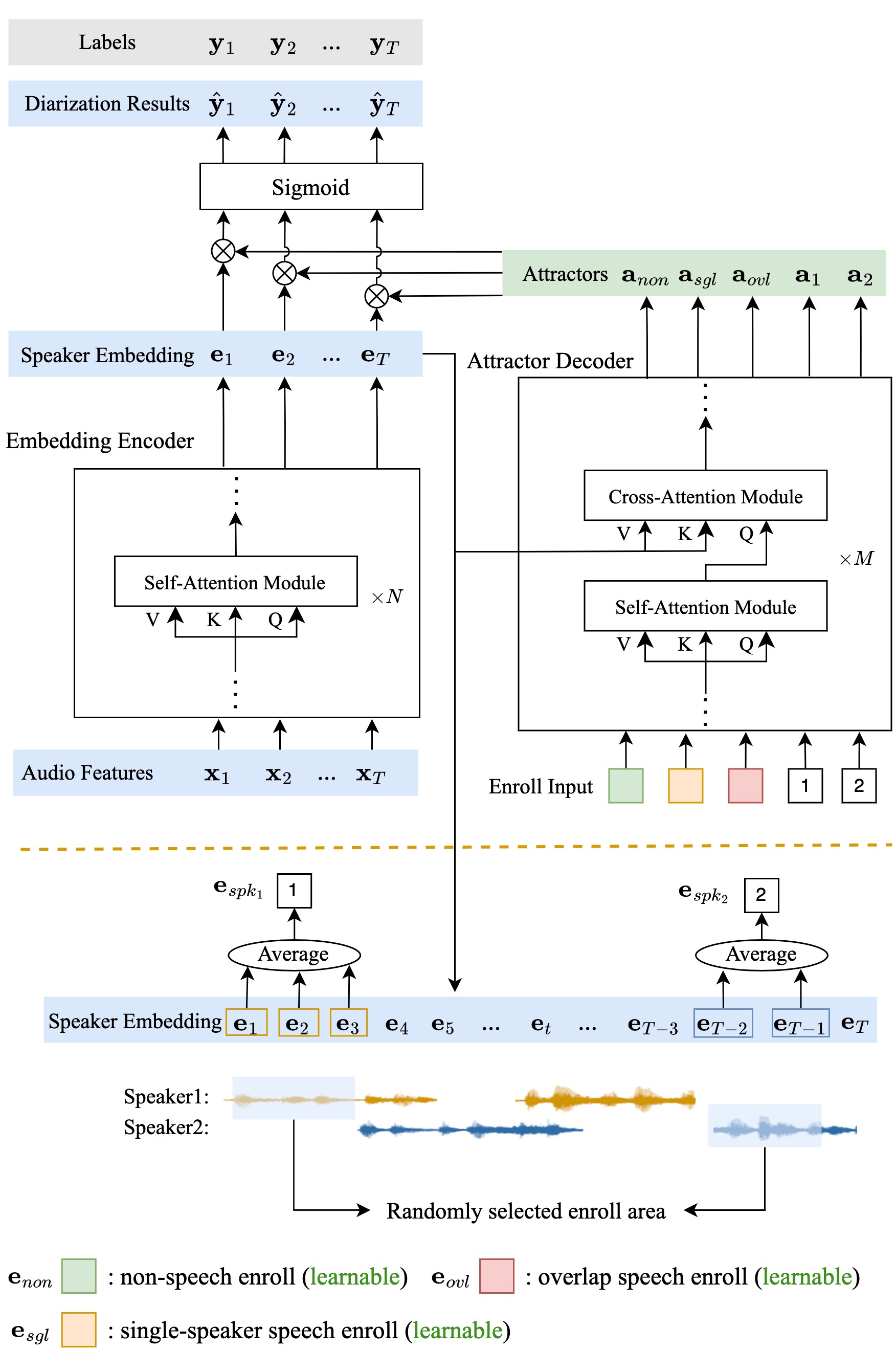}
  \caption{\textbf{AED-EEND system architecture when the speaker number is two.} The part above the orange dotted line is our main system architecture, which is introduced in section \ref{ssec:aed-nd_introduction}. The part below the line shows our strategy to get the enrollment embedding in the training process, which is introduced in \ref{ssec:teacher_forcing_train_intro}. }
  \label{fig:aed-nd}
\end{figure}

\subsection{Attention-based Encoder-Decoder Neural Diarzation}
\label{ssec:aed-nd_introduction}
In this section, we introduce our proposed AED-EEND system. We just follow the architecture of the original transformer encoder-decoder network proposed in \cite{vaswani2017attention}. We made a modification by excluding the use of positional embeddings, similar to the approach taken in \cite{horiguchi2022encoder}, as we found that it did not significantly impact the system's performance. 
As depicted in the upper part of Figure \ref{fig:aed-nd}, the encoder in our AED-EEND system takes the audio feature sequence, denoted as $X = [\mathbf{x}_1, \mathbf{x}_2, ..., \mathbf{x}_T] \in \mathbb{R}^{T \times F}$, as input and produces the frame-level speaker embedding sequence denoted as $E = [\mathbf{e}_1, \mathbf{e}_2, ..., \mathbf{e}_T]  \in \mathbb{R}^{T \times D}$. On the other hand, the decoder, referred to as the attractor decoder, follows the naming convention used in \cite{horiguchi2020end,horiguchi2022encoder}. The attractor decoder takes a sequence of enrollment embeddings, denoted as $E_\text{enroll} = [\mathbf{e}_\text{non}, \mathbf{e}_\text{sgl}, \mathbf{e}_\text{ovl}, \mathbf{e}_{\text{spk}_1}, ..., \mathbf{e}_{\text{spk}_S}]  \in \mathbb{R}^{(S+3) \times D}$, as input. Here, $\mathbf{e}_\text{non}$, $\mathbf{e}_\text{sgl}$ and $\mathbf{e}_\text{ovl}$ represent the enrollment embeddings for non-speech, single-speaker speech, and overlapping speech, respectively. Furthermore, $\mathbf{e}_{\text{spk}_i}$ corresponds to the enrollment embedding for the $i$-th speaker in the recording. The attractor decoder generates the corresponding attractor, denoted as $A = [\mathbf{a}_\text{non}, \mathbf{a}_\text{sgl}, \mathbf{a}_\text{ovl}, \mathbf{a}_{\text{spk}_1}, ..., \mathbf{a}_{\text{spk}_S}]  \in \mathbb{R}^{(S+3) \times D}$, for each enrollment input.

Our decoder architecture in the AED-EEND system differs from the LSTM-based attractor decoder presented in \cite{horiguchi2020end}, as it incorporates both a self-attention module and a cross-attention module. The self-attention module enables the enrollment embeddings to interact with each other, resulting in more distinct attractors being generated. Meanwhile, the cross-attention module allows the enrollment embeddings to attend to all frame-level speaker embeddings, ensuring that the output attractors capture more relevant information from the frame-level speaker embeddings.
In contrast to the EEND-EDA system described in \cite{horiguchi2020end}, which only outputs attractors for existing speakers in the recording, our proposed AED-EEND system also produces attractors for three different speech activities: non-speech, single-speaker speech, and overlapping speech. Originally, this design choice was motivated by the decoding algorithm outlined in section \ref{ssec:decoding_intro}, which necessitates predicting the speech of a single person. Moreover, we believe that modeling a broader range of attractors contributes to improved system performance. Additionally, the results presented in section \ref{sec:speech_activity_res} demonstrate that our AED-EEND system can independently function as a speech type detection system.

With the extracted frame-level speaker embedding sequence $E$ and attractor sequence $A$, we can calculate the posterior probability that each speaker embedding belongs to the specific attractor based on a simple dot product operation:

\begin{equation}
\label{eq:get_posterior}
\hat{Y}=\sigma\left(A E^{\top}\right) \in(0,1)^{(S+3) \times T}
\end{equation}
where the $\sigma(\cdot)$ symbol corresponds to the element-wise sigmoid function. Here, we denote the posterior probabilities for all the speech activities and speakers at the $t^{th}$ frame as: $\hat{\mathbf{y}}_{t} =[\hat{y}_t^{\text{non}}, \hat{y}_t^{\text{sgl}}, \hat{y}_t^{\text{ovl}}, \hat{y}_t^{1}, ..., \hat{y}_t^{S}] \in(0,1)^{(S+3)}$ and we denote the correponding groud-truth label at $t^{th}$ frame as $\mathbf{y}_{t} =[y_t^{\text{non}}, y_t^{\text{sgl}}, y_t^{\text{ovl}}, y_t^{1}, ..., y_t^{S}] \in\{0,1\}^{(S+3)}$. A label value of 1 indicates the presence of speech activity or a speaker in the $t^{th}$ frame, while 0 indicates the absence. Then, we calculate the loss for each utterance by averaging the binary cross-entropy between posterior probability and ground-truth label across all the attractors and frames:

\begin{equation}
\mathcal{L}=\frac{1}{T(S+3)}\sum_{t=1}^T\sum_{s\in \mathbb{S}} \left[ -y_t^s \log \hat{y}_t^s-\left(1-y_t^s\right) \log \left(1-\hat{y}_t^s\right)\right]
\end{equation}
where $\mathbb{S} = \{\text{non}, \text{sgl}, \text{ovl}, 1, .., S\}$.

\subsection{Frame-level Speaker Embedding Enhancer}

\begin{figure}[ht!]
  \centering
  \includegraphics[width=0.46\textwidth]{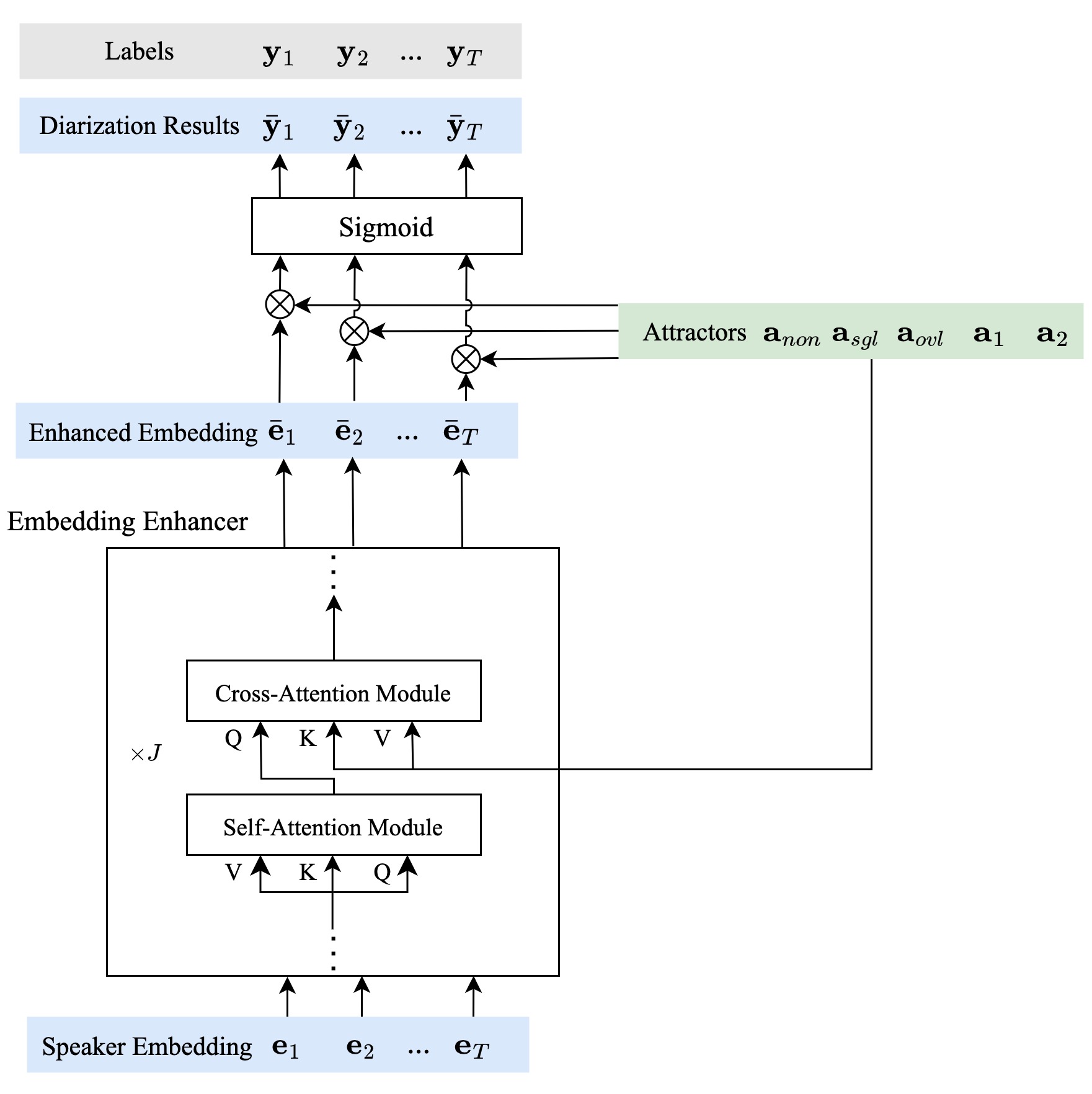}
  \caption{\textbf{The Embedding Enhancer module when the speaker number is two.} }
  \label{fig:enhancer}
\end{figure}

In the cross-attention module of the attractor decoder within our proposed AED-EEND system, we utilize the enrollment input or intermediate outputs as the query, while the frame-level embeddings from the encoder serve as the key and value. This architectural design enables the model to extract valuable information from the frame-level embeddings, leading to the generation of more informative attractors. In a similar vein, we explore the possibility of designing a complementary structure that allows the frame-level embeddings to gather more useful information from the attractors, thereby improving the quality of the frame-level embeddings. To this end, we introduce the Embedding Enhancer (EE) module, which is illustrated in Figure \ref{fig:enhancer}. The EE module takes the frame-level speaker embeddings and all the attractors as input. In contrast to the attractor decoder shown in Figure \ref{fig:aed-nd}, the EE module treats the speaker embeddings as the query in the cross-attention module, while the attractors are used as the key and value. This arrangement facilitates the flow of information from the attractors to the frame-level speaker embeddings. Surprisingly, results in Table \ref{table:flexible_spk_num_callhome_res} and Table \ref{table:dihard_res} indicate that this module can help the system generalize better to evaluation sets with unseen numbers of speakers. Here, we denote the enhanced embedding obtained from the EE module as $\bar{E} = [\mathbf{\bar{e}}_1, \mathbf{\bar{e}}_2, ..., \mathbf{\bar{e}}_T]  \in \mathbb{R}^{T \times D}$. By applying the similar operation described in equation \ref{eq:get_posterior}, we derive the updated posterior probabilities as $\bar{\mathbf{y}}_{t} =[\bar{y}_t^{\text{non}}, \bar{y}_t^{\text{sgl}}, \bar{y}_t^{\text{ovl}}, \bar{y}_t^{1}, ..., \bar{y}_t^{S}] \in(0,1)^{(S+3)}$. The loss function for the enhanced embedding can be defined as follows:

\begin{equation}
\mathcal{L}_{EE}=\frac{1}{T(S+3)}\sum_{t=1}^T\sum_{s\in \mathbb{S}} \left[ -y_t^s \log \bar{y}_t^s-\left(1-y_t^s\right) \log \left(1-\bar{y}_t^s\right)\right]
\end{equation}

When we add the EE module to our AED-EEND system and get the AED-EEND-EE system, we optimize the whole system with the summation of loss $\mathcal{L}$ and loss $\mathcal{L}_{EE}$:
\begin{equation}
\mathcal{L}_{total} = \mathcal{L} + \mathcal{L}_{EE}
\end{equation}

\subsection{Model Optimization with Teacher Forcing Strategy}
\label{ssec:teacher_forcing_train_intro}

As described in Section \ref{ssec:aed-nd_introduction}, our attractor decoder requires enrollment embeddings for each kind of speech activity and all the speakers existing in the utterance as input. This concept of enrollment embedding closely resembles that of the TS-VAD system \cite{medennikov2020target}. In the TS-VAD system, the enrollment embedding is obtained from a pre-trained speaker embedding extractor. In Figure \ref{fig:aed-nd}, the bottom part illustrates that we derive the speaker enrollment embedding directly from the frame-level speaker embedding sequence by averaging a subset of embeddings associated with the specific speaker. Moreover, unlike the speakers' identities, which vary across utterances, the three types of speech activities exist in all utterances. Hence, we directly set the enrollment embeddings for the three distinct types of speech activity as learnable vectors.

In the training of automatic speech recognition (ASR) systems, the teacher forcing strategy \cite{woodward2020confidence} is commonly employed. This strategy involves feeding the system with the ground-truth word (token) to predict the subsequent word (token). By adopting this approach, model training becomes more stable, and convergence is achieved at a faster rate. Inspired by a similar concept, we adopt a comparable strategy in our training phase. We extract the single-speaker speaking region for each individual from the ground-truth label and randomly sample a contiguous region with a predefined enrollment length (EL), denoted as $L_{\text{Enroll}}$, as the enrollment area. Subsequently, we compute the enrollment embedding by averaging the speaker embeddings within the enrollment area.

\begin{figure}[ht!]
  \centering
  \includegraphics[width=0.46\textwidth]{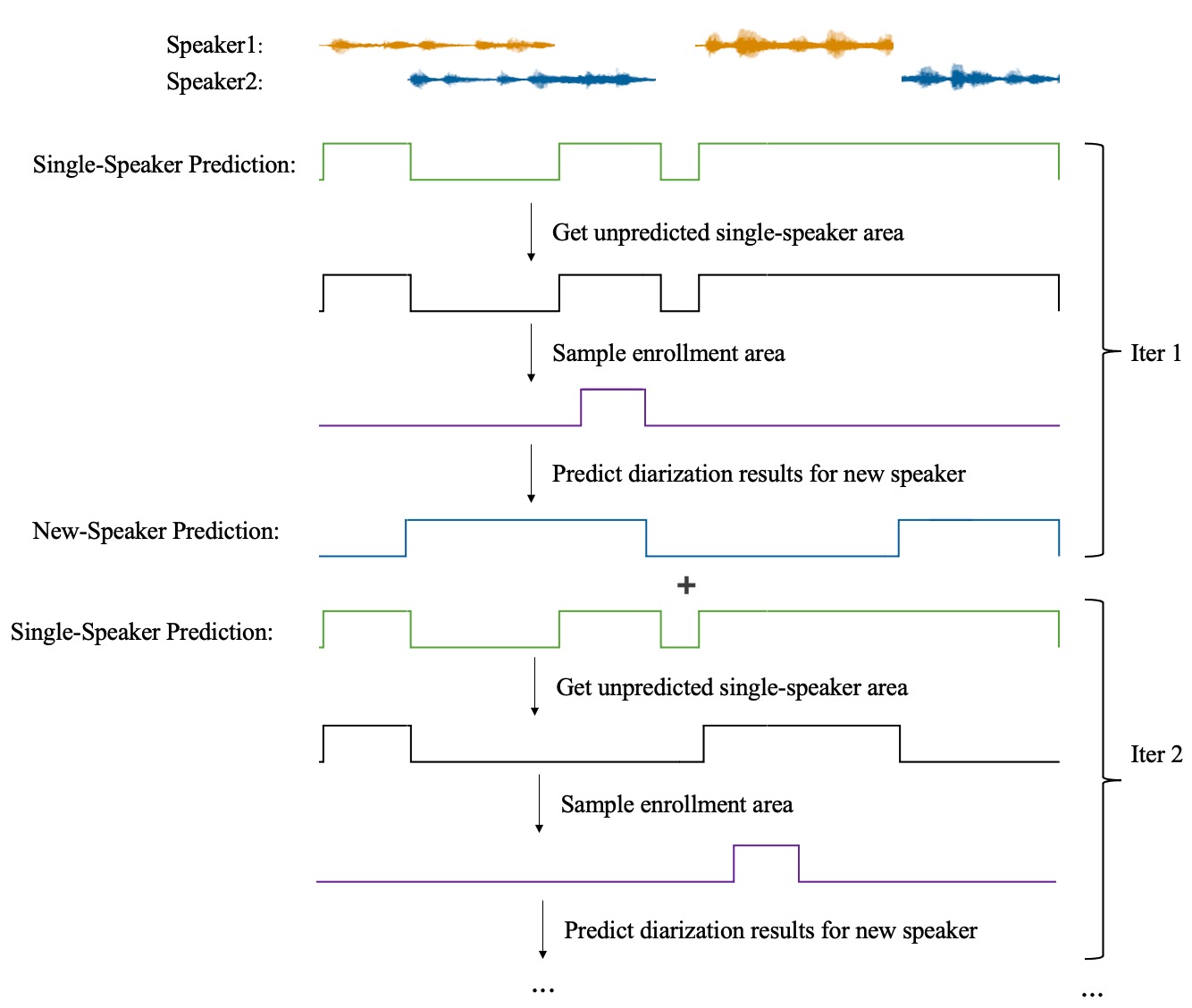}
  \caption{\textbf{The overall pipeline of the iterative decoding method.} }
  \label{fig:decoding_pipeline}
\end{figure}

\subsection{Iterative Decoding Algorithm for Inference}
\label{ssec:decoding_intro}

In this section, we present our proposed iterative decoding method for the AED-EEND system. As discussed in section \ref{ssec:teacher_forcing_train_intro}, we utilize the ground-truth label to find the enrollment area for each speaker in the training process, but such a strategy can not be used in the evaluation process. In the TS-VAD method \cite{medennikov2020target}, an extra diarization system is used to get the single-speaker speaking area for each person. We can certainly use the same method as TS-VAD to obtain the enrollment area, but for the simplicity of the system, we propose an iterative decoding method based on some heuristic strategies to find the single-speaker speaking area for each speaker.

The proposed iterative decoding method follows a general process, as illustrated in Figure \ref{fig:decoding_pipeline}, which encompasses three key operations:
\begin{enumerate}
    \item Get the unpredicted single-speaker speaking area.
    \item Sample a continuous area from the unpredicted single-speaker speaking area, while imposing constraints to ensure that only one speaker is present within the chosen area.
    \item Use the sampled continuous area as the new enrollment area to predict the diarization results for a new speaker.
    \item Repeat steps one to three until the results of all speakers have been decoded.
\end{enumerate}

\begin{algorithm}[!ht]
\footnotesize

\KwData{
$I = [t_1, t_2, ...]$: frame indexes list for embedding in $E$ \\
\hspace{20pt} $C = [I_1, I_2, ..., I_K]$: $K$ time-continuous segments\\
\hspace{24pt} from $I$, and the indexes in $I_{k}$ are sorted in order\\
\hspace{20pt} Enroll Length (EL): $L_{\text{Enroll}}$ \\
\hspace{20pt} Stop Decoding Length (SDL): $L_{\text{stop}}$ \\




}

 \tcp{get the active frame indexes list for three kinds of speeches}
 $I_{\text{non}}, I_{\text{sgl}}, I_{\text{ovl}}  = \text{AED-EEND}(\mathbf{e}_\text{non}, \mathbf{e}_\text{sgl}, \mathbf{e}_\text{ovl})$ \\
 \tcp{active frame indexes list for speakers; enroll embedding set}
 $I_{\text{spk}}=[]; \mathcal{E}=\{\}$

 \While{True}{
        \tcp{get the un-predicted single-speaker area}
        $I = I_{\text{sgl}} - I_{\text{sgl}} \cap I_{\text{spk}}$ \\

        \tcp{get segments with length $\geq L_{\text{Enroll}}$}
        $C' = [I'_1, I'_2, ..., I'_{K'}] = \text{filter\_segs}(C, L_{\text{Enroll}})$ \\
        \tcp{get the segment with the longest length}
            $I_{\text{longest}} = \text{get\_longest\_seg}(C')$ \\
        $L_{\text{Enroll\_tmp}} = \text{min}(\text{length}(I_{\text{longest}}), L_{\text{Enroll}})$ \\
        
        \If{$\text{length}(C') == 0)$}
            {$C'\text{.add}(I_{\text{longest}})$} 
            
        \If{$L_{\text{\text{stop}}} > \text{length}(I_{\text{longest}})$}
            {break; \tcp{stop decoding}}

        \Case{\textbf{Init-Decode}}{
            
            $I_{\text{enroll}} = [t_1, t_2, ..., t_{L_{\text{Enroll\_tmp}}}] \in I'_{1}$
            
        }
        \Case{\textbf{Rand-Decode}}{
            \tcp{randomly select a segment from $C'$}
            $I'_k = \text{random\_select\_seg}(C')$ \\
            \tcp{randomly select a consecutive sub-segment with length $L_{\text{Enroll\_tmp}}$}
            $I_{\text{enroll}} = \text{random\_select\_sub-seg}(I'_k, L_{\text{Enroll\_tmp}})$ \\
        
        }
        \Case{\textbf{SC-Decode}}{
            \tcp{The index in $I$ is clustered based on corresponding embedding}
            $[I^{\text{cls}}_1, I^{\text{cls}}_2, ...] = \text{spectral\_cluster}(I)$ \\
            $I^{\text{cls}}_k = \text{get\_longest\_seg}([I^{\text{cls}}_1, I^{\text{cls}}_2, ...])$ \\
            $I_{\text{enroll}} = \text{random\_select\_sub-seg}(I^{\text{cls}}_k, L_{\text{Enroll\_tmp}})$ \\
            
        }
        \Case{\textbf{SC-Decode-Local}}{
            $[I^{\text{cls}}_1, I^{\text{cls}}_2, ...] = \text{spectral\_cluster}(I_{\text{longest}})$ \\
            $I^{\text{cls}}_k = \text{get\_longest\_seg}([I^{\text{cls}}_1, I^{\text{cls}}_2, ...])$ \\
            $I_{\text{enroll}} = \text{random\_select\_sub-seg}(I^{\text{cls}}_k, L_{\text{Enroll\_tmp}})$ \\
            
        }
        \tcp{average the embeddings with indexes in $I_{\text{enroll}}$}
        $\mathbf{e} = \text{average}(E_{I_{\text{enroll}}})$\\
        $\mathcal{E}\text{.add}(\mathbf{e})$\\ 
        $I_{\text{spk}} = (\text{AED-EEND}(\mathcal{E}))$

}
\KwOut{$I_{\text{spk}}$}

\caption{Proposed Iterative Decoding Algorithm}
\label{algo:decoding}
\end{algorithm}

The detailed implementation of our iterative decoding method is outlined in Algorithm \ref{algo:decoding}. Initially, we input the enrollment embeddings $[\mathbf{e}_\text{non}, \mathbf{e}_\text{sgl}, \mathbf{e}_\text{ovl}]$ into our AED-EEND system to obtain the index list $I_{\text{sgl}}$  representing frames with single-speaker activity. Subsequently, we derive the index list $I$ containing the frames that exist in $I_{\text{sgl}}$ but not in $I_{\text{spk}}$, where $I_{\text{spk}}$ encompasses the frame indices predicted for the speakers in previous iterations. The un-predicted single-speaker area is thus captured by $I$.
Next, our aim is to find a continuous area of enrollment length (EL) $L_{\text{Enroll}}$ from $I$ to serve as the new enrollment area, with the objective of ensuring that this area contains only one speaker. To accomplish this, we have devised four heuristic strategies:

\begin{itemize}
    \item \textbf{Init-Decode}: In this strategy, we sample the initial $L_{\text{Enroll}}$ length area from the first continuous segment in $I$ as the new enrollment area.
    \item \textbf{Random-Decode}: Here, we randomly select a continuous segment, denoted as $I'_k$, from $I$. From $I'_k$, we further randomly select a continuous area with a length of $L_{\text{Enroll}}$ as the new enrollment area
    \item \textbf{SC-Decode}: This strategy involves performing spectral clustering \cite{wang2018speaker,lin2019lstm} on all the embeddings in $I$ to obtain the embedding index cluster list $I^{\text{cls}}_k$ with the longest length. From $I^{\text{cls}}_k$, we randomly sample a segment with a length of $L_{\text{Enroll}}$ as the new enrollment area.
    \item \textbf{SC-Decode-Local}: We observed that spectral clustering can be time-consuming for long sequences. In this strategy, we limit the spectral clustering process to the longest continuous segment within $I$. Subsequently, we apply the same procedure as SC-Decode to identify the new enrollment area. It is worth noting that, because this method only performs clustering locally, it is much faster than the SC-Decode.
\end{itemize}

While the Init-Decode and Random-Decode strategies do not explicitly impose constraints to ensure the selection of areas with only one speaker, the experiments conducted in section \ref{sssec:decoding_method_compare} demonstrate that using a very short enrollment area can yield remarkably good performance. The utilization of a short enrollment area significantly increases the likelihood of capturing an area containing only one speaker. Furthermore, we have developed a criterion to determine when to terminate the iterative decoding process. Specifically, when the length of the longest continuous segment within $I$ is below a pre-defined stop decoding length (SDL) of $L_{\text{stop}}$, the decoding process is terminated.

\section{Experimental Setup}

\subsection{Data Corpus}
\subsubsection{Dataset with Simulation Recordings}
\label{sssec:simu_data_intro}
To have a fair comparison with the EEND-EDA system, we just follow \cite{horiguchi2022encoder} to simulate 5 different sub-sets of simulation data. The detailed information regarding the simulation data can be found in the upper part of Table \ref{table:simulation_dataset}.

\begin{table}[ht!]
\footnotesize
\centering
\caption{\textbf{Simulation Dataset Configuration}.  Ovl corresponds to the overlap ratio.}

\begin{adjustbox}{width=.49\textwidth,center}
\begin{threeparttable}
\begin{tabular}{lcccccc}
\toprule Dataset & Split & \#Spk & \#Mixtures & $\beta$ & Ovl $(\%)$ & Duration (hrs) \\
\hline SM-1spk & Train & 1 & 100,000 & 2 & 0.0 & 2159\\
& Test & 1 & 500 & 2 & 0.0 & 10.76 \\
\hline SM-2spk & Train & 2 & 100,000 & 2 & 34.1 & 2484 \\
& Test & 2 & 500 & 2 & 34.4 & 12.33 \\

\hline SM-3spk & Train & 3 & 100,000 & 5 & 34.2 & 4226 \\
& Test & 3 & 500 & 5 & 34.7 & 20.92 \\

\hline SM-4spk & Train & 4 & 100,000 & 9 & 31.5 & 6647 \\
& Test & 4 & 500 & 9 & 32.0 & 33.03\\
\hline SM-5spk & Train & 5 & 100,000 & 13 & 30.3 & 9202 \\
& Test & 5 & 500 & 13 & 30.7 & 45.20\\
\midrule
\midrule
SC-1spk & Train & 1 & 36,989 & - & 0.0 & 2159\\
& Test & 1 & 223 & - & 0.0 & 12.45 \\
\hline SC-2spk & Train & 2 & 24,343 & - & 8.18 & 2481 \\
& Test & 2 & 118 & - & 8.13 & 12.41 \\

\hline SC-3spk & Train & 3 & 29,297 & - & 11.5 & 4226 \\
& Test & 3 & 86 & - & 11.1 & 12.44 \\

\hline SC-4spk & Train & 4 & 35,640 & - & 13.4 & 6647 \\
& Test & 4 & 66 & - & 12.9 & 12.43\\
\hline SC-5spk & Train & 5 & 40,249 & - & 14.5 & 9202 \\
& Test & 5 & 55 & - & 14.8 & 12.50\\

\bottomrule
\end{tabular}

\end{threeparttable}
\label{table:simulation_dataset}
\end{adjustbox}
\end{table}

The utterances in each subset have a fixed number of speakers, and the number of speakers varies across different subsets, ranging from 1 to 5. Although this simulation setup has been widely used in previous works \cite{fujita2019end_lstm,fujita2019end_sa,fujita2020neural,horiguchi2020end,horiguchi2022encoder}, it should be noted that there is a significant mismatch in the overlap ratio between the simulated data and the real recordings, as indicated in Table \ref{table:real_recoding_dataset}. We believe that this configuration may not be optimal. 
In recent studies \cite{landini22_interspeech,landini2023multi}, the authors proposed a methodology that leverages the statistics from real recordings to guide the synthesis of simulated data, resulting in improved performance on real datasets. Additionally, the authors in \cite{landini22_interspeech} referred to the simulation data in \cite{horiguchi2020end,horiguchi2022encoder} as simulated mixtures (SM), while they named the simulation data in their own paper as simulated conversion (SC). In our paper, we adopt these terminologies. Following the approach outlined in \cite{landini22_interspeech}, we extracted statistics from Part1 of the CALLHOME dataset (as shown in Table \ref{table:real_recoding_dataset}) and simulated an equal amount of SC data to match the SM data. The statistics presented in Table \ref{table:simulation_dataset} illustrate that the SC data exhibits a significantly smaller overlap ratio compared to the SM data. 
While the effectiveness of SC data has been validated in previous studies \cite{landini22_interspeech,landini2023multi}, it is important to note that the experiments in \cite{landini22_interspeech} were primarily conducted using only the 2-speaker CALLHOME data and DIHARD III CTS data, and the baselines employed in \cite{landini2023multi} were relatively weaker. In our experiments, we aim to evaluate the effectiveness of SC data in more diverse scenarios and employ a strong system.

\begin{table}[ht!]
\footnotesize
\centering
\caption{\textbf{Datasets of Real Recordings}. The three numbers in the Duration column correspond to the minimum duration/maximum duration/average duration. Ovl corresponds to the overlap ratio. }

\begin{adjustbox}{width=.49\textwidth,center}
\begin{threeparttable}
\begin{tabular}{lccccc}
\toprule Dataset & Split & \#Spk & \#Utt & Ovl (\%) & Duration (mins)\\
\hline \multirow{2}{*}{CALLHOME-2spk \cite{callhome}} & Part 1 & 2 & 155 & 14.0 & 0.862/2.211/1.234 \\
& Part 2 & 2 & 148 & 13.1 & 0.875/2.233/1.202 \\
\hline \multirow{2}{*}{CALLHOME-3spk \cite{callhome}} & Part 1 & 3 & 61 & 19.6 & 0.947/6.352/2.071 \\
& Part 2 & 3 & 74 & 17.0 & 0.774/8.210/2.418 \\
\hline \multirow{2}{*}{CALLHOME \cite{callhome}} & Part 1 & $2-7$ & 249 & 17.0 & 0.862/10.11/2.097 \\
& Part 2 & $2-6$ & 250 & 16.7 & 0.774/10.01/2.053 \\
\hline \multirow{3}{*}{AMI headset mix \cite{carletta2006ami}} & Train & $3-5$ & 136 & 13.4 & 7.965/90.25/35.59 \\
& Dev & 4 & 18 & 14.1 & 15.73/49.50/32.22  \\
& Test & $3-4$ & 16 & 14.6 & 13.98/49.53/33.98 \\
\hline \multirow{2}{*}{DIHARD II \cite{dihard2}} & Dev & $1-10$ & 192 & 9.8 & 0.447/11.62/7.442 \\
& Test & $1-9$ & 194 & 8.9 & 0.631/13.50/6.957 \\
\bottomrule
\end{tabular}

\end{threeparttable}
\label{table:real_recoding_dataset}
\end{adjustbox}
\end{table}

\subsubsection{Dataset with Real Recordings}
\label{sssec:real_recoding_dataset}

In our experiments, we evaluate our systems on three different datasets with real recordings, the CALLHOME dataset \cite{callhome}, AMI dataset with mixed headset \cite{carletta2006ami} and the DIHARD II \cite{dihard2} dataset \footnote{Unfortunately, we do not report results for the DIHARD III dataset as it is not freely available, and we only have access to the DIHARD II dataset.}. The CALLHOME dataset is divided into two parts according to the kaldi recipe \footnote{\url{https://github.com/kaldi-asr/kaldi/tree/master/egs/callhome_diarization/v2}}. Part 1 is used for model adaptation, while Part 2 is used for evaluation. The AMI dataset consists of three parts: Train, Dev, and Test. We perform model adaptation on the Train part and evaluate the system on both the Dev and Test parts. For the DIHARD II dataset, we conduct model adaptation on the Dev part and evaluate on the Test part. The CALLHOME dataset comprises 8k telephone-channel recordings, the AMI dataset consists of 16k meeting recordings, and the DIHARD II dataset contains 16k recordings from a diverse range of sources. To simplify the training setup, we downsampled the AMI and DIHARD II datasets to 8k in our experiments.

\subsection{Model Configuration}
\label{ssec:model_configuration_intro}

For all the audio data in our experiments, we extract 345-dimensional acoustic features following the methodology described in \cite{horiguchi2020end,horiguchi2022encoder}. Consequently, the input dimension for all our systems is set to 345. In the attention module, we configure the number of heads to 4 and the attention unit number to 256. Additional system configurations can be found in Table \ref{table:model_config}. It should be noted that, when we use the Embedding Enhancer (EE) module, we share parameters between the second to fourth layers of the decoder and Enhancer, which equals that the decoder and Enhancer only have two layers containing parameters. With this setup, our AED-EEND-EE system has the same number of parameters as the AED-EEND system. 
Additionally, we experimented with replacing the transformer encoder with the Conformer \cite{gulati2020conformer} module. The Conformer module uses a convolution kernel size of 31, and we adjusted the feed-forward dimension of the encoder and decoder to achieve a similar parameter count as the AED-EEND system. Although the AED-EEND system shares the same encoder as the EEND-EDA system, the attention-based decoder in the AED-EEND system has more parameters compared to the LSTM-based decoder in the EEND-EDA system. To ensure a fair parameter comparison, we designed a smaller system called AED-EEND-EE (small), which matches the parameter count of the EEND-EDA system.

\begin{table}[ht!]
\footnotesize
\centering
\caption{\textbf{Model Configuration}.}

\begin{adjustbox}{width=.45\textwidth,center}
\begin{threeparttable}
\begin{tabular}{lcccc}

\toprule
\multirow{2}{*}{Model} & \multirow{2}{*}{Param \#} &  \multicolumn{3}{c}{Layer\# / Feed-Forward Dimension}  \\
\cmidrule(r){3-5}
 & &  Encoder & Decoder & Enhancer \\
\hline
AED-EEND & 11.6M  & 4/2048 & 4/2048 & 0/- \\
AED-EEND (conformer) &  10.4 M & 4/1024 & 4/1024 & 0/- \\
AED-EEND-EE & 11.6M & 4/2048 & 4/2048 & 4/2048 \\
AED-EEND-EE (small) & 6.4M & 4/1024 & 4/512 & 4/512 \\
AED-EEND-EE (conformer) & 10.4M & 4/1024 & 4/1024 & 4/1024 \\
\hline
EEND-EDA & 6.4M & 4/2048 & - & - \\

\bottomrule
\end{tabular}

\end{threeparttable}
\label{table:model_config}
\end{adjustbox}
\end{table}

\subsection{Training Setup}
\label{ssec:training_setup}

In our experiment, the training process is divided into two stages: pre-training with the simulated dataset and adaptation with the real dataset. The model trained solely on the simulation data is evaluated on the simulation evaluation set, while the adapted model is evaluated on the real dataset. We randomly sample the enrollment length introduced in section \ref{ssec:teacher_forcing_train_intro} from 1s to 3s. We also randomly drop the enrollment embeddings for all the speakers existing in each utterance with a probability of 0.5 in the training process, which ensures the system's robustness when encountering scenarios where not all speakers' enrollment embeddings are available during the iterative decoding process.

The acoustic features mentioned in section \ref{ssec:model_configuration_intro}, with a dimensionality of 345, are extracted using a window size of 25ms and a hop length of 10ms. During the training process, we downsample the acoustic feature sequence by a factor of ten, resulting in a frame resolution of 0.1s. 
For the pre-training stage, all utterances are divided into segments of 50s in length, from which we randomly sample 64 segments to construct the training batch. As for the adaptation stage, following the approach in \cite{horiguchi2022encoder}, we use 50s segments for the CALLHOME dataset and 200s segments for the AMI and DIHARD II datasets. In the adaptation stage, we set the batch size to 32. 
During the pre-training stage, we utilize the Noam optimizer \cite{vaswani2017attention} and configure the training process with 100 epochs and 200,000 warmup steps. For the adaptation stage, we employ the Adam optimizer with a learning rate of 0.00001.

In our experiment, the evaluation of the system can be categorized into two scenarios: the fixed number-of-speaker scenario and the flexible number-of-speaker scenario.
In the fixed number-of-speaker scenario, the pre-training data, adaptation data, and evaluation data have the same number-of-speaker. In the flexible number-of-speaker scenario, we perform pre-training by combining the simulation data from Table \ref{table:simulation_dataset} for a range of speaker numbers, either from 1 speaker to 4 speakers or from 1 speaker to 5 speakers, depending on the specific experiment. 
Subsequently, we adapt the pre-trained model using the specific real adaptation set.

\begin{table*}[ht!]
\footnotesize
\centering
\caption{\textbf{DER (\%) results comparison for different decoding methods with varying enroll lengths and stop decoding lengths.} 
All the results are based on our 2-spk systems.
\textbf{EL}: enroll length. \textbf{SDL}: stop decoding length. \textbf{GT-Decode} stands for ground truth decode, which refers to the teacher-forcing strategy employed during the training process. Results marked with a \colorbox{lightgray}{gray background} indicate very poor performance. In the results in the upper part of the table, we assume that the number of speakers is known, which is 2. In the results of the lower part of the table, the number of speakers is determined through SDL.
}
\begin{adjustbox}{width=.98\textwidth,center}
\begin{tabular}{c ccccccc ccccccc}
\toprule
\multirow{2}{*}{Decoding Method} & \multicolumn{7}{c}{DER (\%) on 2-spk SM with Different EL} & \multicolumn{7}{c}{DER (\%) on 2-spk CALLHOME with Different EL}\\

\cmidrule(r){2-8} \cmidrule(r){9-15} 

& 0.1s & 0.5s & 1s & 2s & 3s & 5s & 10s & 0.1s &  0.5s & 1s & 2s & 3s & 5s & 10s \\
\hline

GT-Decode & 3.10 & 3.22 & 3.22 & 2.95 & 3.05  & 2.87  & 2.88  & 8.39 & 8.03  & 7.58 & 7.56 & 7.18  & 7.27  & 7.19  \\
Init-Decode & 5.18 & 3.14 & 3.14 & 6.34 & \cellcolor{lightgray} 23.8 & \cellcolor{lightgray} 67.1 & \cellcolor{lightgray} 97.5 & \cellcolor{lightgray} 40.78 & 10.9 & 7.90 & 8.71 & 15.1 & \cellcolor{lightgray} 35.1 & \cellcolor{lightgray} 86.1 \\
Rand-Decode & 3.36 & 3.13 & 3.18 & 6.73 & \cellcolor{lightgray} 24.1 & \cellcolor{lightgray} 67.2 & \cellcolor{lightgray} 97.5 & 11.36 & 8.32  & 8.38 & 9.39 & 16.0 & \cellcolor{lightgray} 35.5 & \cellcolor{lightgray} 86.1 \\
SC-Decode & 3.13 & 3.14 & 3.12 & 3.17 & 3.16  & 3.17  & 8.27  & 8.61 & 7.75  & 7.86 & 8.04 & 7.68  & 7.81  & 8.78 \\
SC-Decode-Local & 3.47 & 3.48 & 3.50 & 6.94 & \cellcolor{lightgray} 24.63  & \cellcolor{lightgray} 67.36  & \cellcolor{lightgray} 97.50  & 8.39 & 7.81  & 7.76 & 8.93 & 15.95  & \cellcolor{lightgray} 35.48  & \cellcolor{lightgray} 86.15 \\


\midrule
 & \multicolumn{7}{c}{DER (\%) on 2-spk SM with Different SDL } & \multicolumn{7}{c}{DER (\%) on 2-spk CALLHOME with Different SDL}\\
\cmidrule(r){2-8} \cmidrule(r){9-15} 
& 0.5s & 0.8s & 1s & 1.5s & 2s & 2.5s & - & 0.5s & 0.8s & 1s & 1.5s & 2s & 2.5s & - \\
\hline
Init-Decode & 4.92 & 3.55 & 3.40 & 3.56 & 4.88 & 10.03 & - & 19.4 & 10.9 & 10.7 & 10.8 & 10.6 & 12.2 & - \\
Rand-Decode & 4.06 & 3.68 & 3.50 & 3.51 & 5.53 & 10.85 & - & 13.2 & 9.35  & 8.58  & 8.20  & 8.58  & 10.4 & - \\
SC-Decode & 3.98 & 3.55 & 3.25 & 3.53 & 6.59 & 13.08 & - & 13.0 & 8.28  & 8.34  & 8.47  & 9.36  & 11.5 & - \\
SC-Decode-Local & 4.35 & 3.71 & 3.57 & 3.61 & 6.75 & 13.80 & - & 11.05 & 7.98  & 7.87  & 8.24  & 9.13  & 11.49 & - \\
\bottomrule
\end{tabular}
\label{table:enroll_thres_len_ablation}
\end{adjustbox}
\end{table*}

\subsection{Evaluation Setup}
\label{ssec:evaluation_setup}
To ensure a fair comparison with the EEND-EDA system \cite{horiguchi2022encoder}, we adopted most of their evaluation configurations. For the CALLHOME dataset, we downsampled the acoustic feature sequence by 10 times and evaluated the DER (Diarization Error Rate) with a collar tolerance of 0.25s. In their study, the authors downsampled the feature sequence by 5 times for the AMI and DIHARD II evaluations, without using collar tolerance. However, in our experiments, we found that the AMI dataset is not highly sensitive to the downsampling rate and contains significantly longer recordings. As a result, we used a downsampling rate of 10 times for the AMI evaluation and 5 times for the DIHARD II evaluation. No collar is used for AMI and DIHARD II evaluation. We use 0.5 as the threshold to get the decision for diarization results. Additionally, it is worth noting that our experiments did not employ any oracle speech segments. This deliberate choice allowed us to assess the performance of our single system independently and evaluate its capabilities in the diarization task without relying on additional information.

\section{Evaluation on Speaker Diarization Task}
\subsection{Fixed Number of Speakers Scenario}

\subsubsection{Decoding Methods Comparison}
\label{sssec:decoding_method_compare}

In Section \ref{ssec:decoding_intro}, we introduced four decoding methods that are utilized for generating the diarization results iteratively. Additionally, we defined two crucial hyperparameters, namely, the enrollment length (EL) and stop decoding length (SDL). These hyperparameters play a significant role in assisting the algorithm to identify the enrollment area and determine the appropriate time to stop the iteration process. In this section, we will provide a comprehensive comparison of the different decoding methods and explore the impact of varying hyperparameter values on the results. Table \ref{table:enroll_thres_len_ablation} presents the corresponding results obtained from the SM simulation dataset and CALLHOME dataset when the number of speakers is 2.

The upper part of Table \ref{table:enroll_thres_len_ablation} investigates the influence of different enrollment length values and the oracle speaker number is used. When utilizing a short enrollment length with the SM dataset, we observe that all of our proposed decoding methods demonstrate reasonable performance, with some approaches even approaching the performance of the GT-Decode method. Because there is a great deal of randomness in Init-Decode and Rand-Decode, the performance degrades seriously when enrollment length becomes longer, which may cause the enrollment part to contain more than one speaker.
Besides, the SC-Decode-Local method also demonstrates poorer performance with longer enrollment lengths.
Similar patterns are observed when evaluating different decoding methods on the CALLHOME dataset. Additionally, the Init-Decode method performs poorly when an extremely short enrollment area (0.1s) is used for the CALLHOME dataset, possibly due to inaccurate boundary predictions. Considering the robustness of a short enrollment area, we adopt a default enrollment length setup of 0.5s for subsequent experiments.

In the subsequent analysis, we depart from the assumption of knowing the oracle number of speakers and instead utilizing a pre-defined stop decoding length (SDL) to determine the number of speakers. The corresponding results are presented in the lower part of Table \ref{table:enroll_thres_len_ablation}. Our findings indicate that employing excessively long or short SDL values yields less desirable results. Remarkably, the most robust outcomes were achieved when utilizing a SDL of 1 second. This observation aligns with intuitive expectations, as an overly long SDL may overlook certain speakers, while an excessively short SDL can lead the system to falsely predict additional speakers. Moving forward, we will adopt a default SDL of 1 second in our subsequent experiments when the speaker number is unknown. Additionally, for scenarios with a fixed number of speakers, we will use the SC-Decode strategy as the default decoding method. To make decoding faster, we will employ the SC-Decode-Local method for scenarios with a flexible number of speakers.

\subsubsection{Comparison between Different Systems}

\begin{table}[ht!]
\footnotesize
\centering
\caption{\textbf{DER (\%) results on the fixed number of speakers condition.}
}
\begin{adjustbox}{width=.47\textwidth,center}
\begin{threeparttable}
\begin{tabular}{lcccc}
\toprule
\multirow{2}{*}{Method} & \multicolumn{2}{c}{SM data DER(\%)}  & \multicolumn{2}{c}{\textsc{CallHome} DER(\%)}\\
\cmidrule(r){2-3} \cmidrule(r){4-5} 
 & 2-spk & 3-spk & 2-spk & 3-spk \\
 \hline
 x-vector clustering \cite{horiguchi2020end}  &  28.77 & 31.78 & 11.53 & 19.01 \\
 BLSTM-EEND \cite{fujita2019end_lstm} & 12.28 & - & 26.03  & - \\
 SA-EEND  \cite{fujita2019end_sa,horiguchi2020end}  & 4.56   & 8.69  & 9.54  & 14.00 \\
 SC-EEND \cite{fujita2020neural} & -   & -  & 8.86  & - \\
 EEND-EDA  \cite{horiguchi2020end} & 2.69   & 8.38  & 8.07  & 13.92 \\
 EEND-EDA  $^\dagger$ & 4.20   & 10.51  & 8.32  & 17.07 \\
 TS-VAD \cite{cheng2022multi} & - & - & 9.51 & - \\
 MTFAD \cite{cheng2022multi} & - & - & 7.82 & - \\
 AED-EEND                   & 3.14  & 5.16 & \textbf{7.75}  & 12.87  \\
 AED-EEND-EE & \textbf{2.46}  & \textbf{4.26} & 8.18  & \textbf{12.21}  \\

\bottomrule
\end{tabular}
\begin{tablenotes}\footnotesize
\item $^\dagger$: our implementation
\end{tablenotes}
\end{threeparttable}
\label{table:fixed_spk_num_res}
\end{adjustbox}
\end{table}
In this section, we compare our proposed system with others when the speaker number is fixed and known in advance. The results in Table \ref{table:fixed_spk_num_res} show that our system achieves the best results on all conditions. Besides, equipped with our proposed Enhancer module, most of the evaluation sets are improved.





\begin{table}[ht!]
\footnotesize
\centering
\caption{\textbf{DER (\%) results on the SM simuation dataset with flexible number of speakers.} Systems are trained on 1-4 speakers SM dataset.
}
\begin{adjustbox}{width=.36\textwidth,center}
\begin{tabular}{lcccc}
\toprule
\multirow{2}{*}{Method} & \multicolumn{4}{c}{spk\#}\\
\cmidrule(r){2-5}
 & 1 & 2 & 3 & 4 \\
 \hline
 x-vector clustering \cite{horiguchi2020end} & & & \\
 \hspace{3pt} Estimated \#spk & 37.42 & 7.74 & 11.46 & 22.45 \\
 \hspace{3pt} Oracle \#spk & 1.67 & 28.77 & 31.78 & 35.76 \\
 
 SC-EEND \cite{fujita2020neural} & & & \\
 \hspace{3pt} Estimated \#spk & 0.76 & 4.31 & 8.31 & 12.50 \\
 
 EEND-EDA \cite{horiguchi2020end} & & & \\
 \hspace{3pt} Estimated \#spk & 0.39 & 4.33 & 8.94 & 13.76 \\
 \hspace{3pt} Oracle \#spk & 0.16 & 4.26 & 8.63 & 13.31 \\
 

 AED-EEND & & & \\
 \hspace{3pt} Estimated \#spk & 0.07 & 2.72 & 5.56 & 10.07 \\
 \hspace{3pt} Oracle \#spk & 0.07 & 2.93 & 6.00 & 9.72 \\


   AED-EEND-EE & & & \\
 \hspace{3pt} Estimated \#spk & \textbf{0.07} & \textbf{2.45} & \textbf{4.71} & 7.04 \\
 \hspace{3pt} Oracle \#spk & 0.09 & 2.73 & 5.83  & \textbf{6.65} \\


\bottomrule
\end{tabular}
\label{table:flexible_spk_num_simu_res}
\end{adjustbox}
\end{table}

\subsection{Flexible Number of Speakers Scenario}
In this section, all the systems will be evaluated on the dataset that the speaker number is flexible.
\subsubsection{Results on Simulation Dataset}

Firstly, we assess the performance of our system on the SM evaluation set and present the results in Table \ref{table:flexible_spk_num_simu_res}, considering scenarios where the speaker number is estimated or the oracle speaker number is known. Notably, our systems exhibit the highest performance among all evaluated systems, and the introduction of our proposed Enhancer module notably enhances results for the 3-speaker and 4-speaker evaluation sets. Furthermore, our systems demonstrate consistent performance regardless of whether the speaker number is known or estimated, indicating the effectiveness of our proposed decoding strategy in accurately predicting the number of speakers.

\begin{figure}[ht]
\begin{subfigure}{.24\textwidth}
  \centering
  \includegraphics[width=1.0\linewidth]{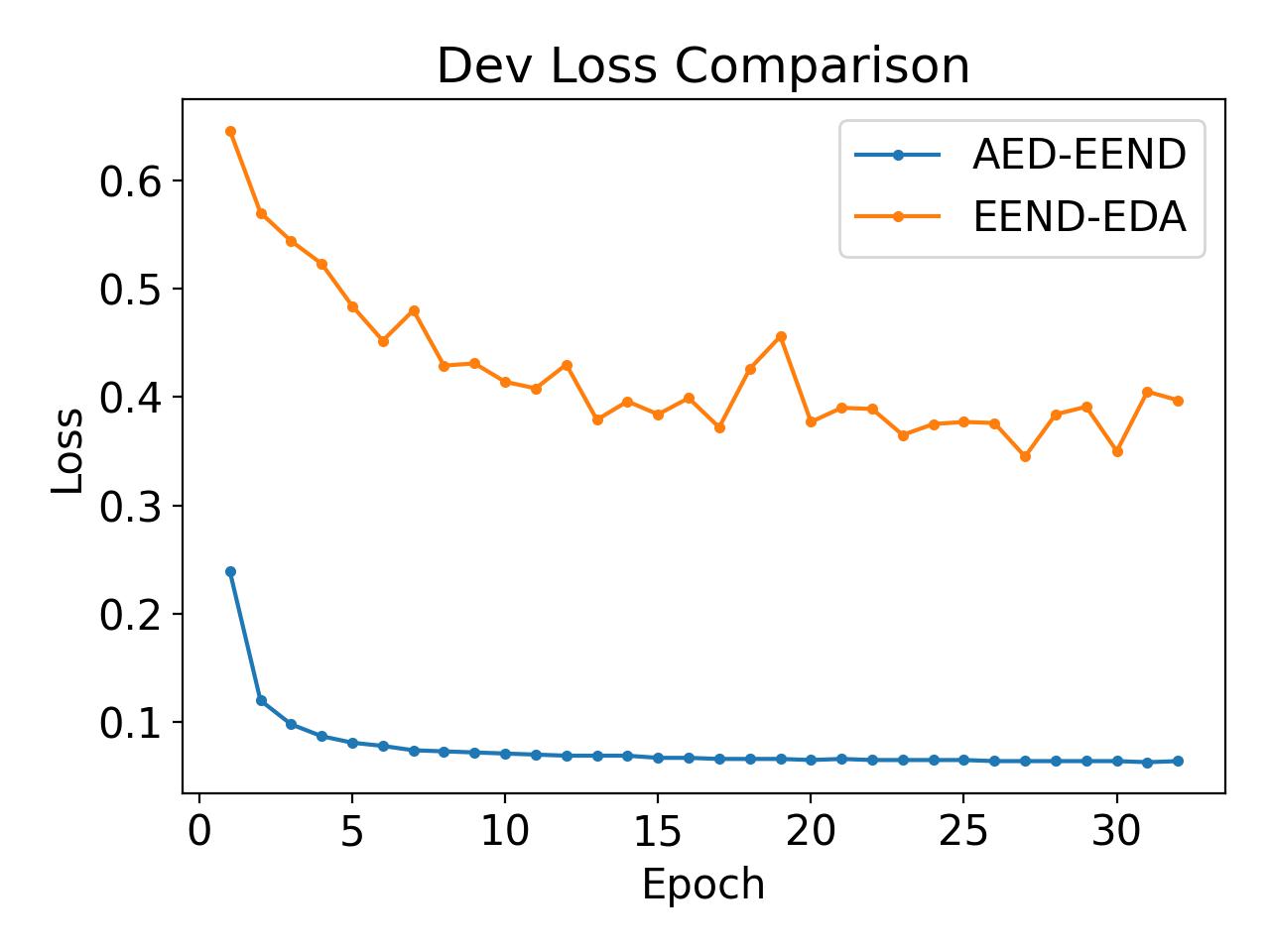}  
  \caption{Loss comparison}
  \label{fig:loss_compare}
\end{subfigure}
\begin{subfigure}{.24\textwidth}
  \centering
  \includegraphics[width=1.0\linewidth]{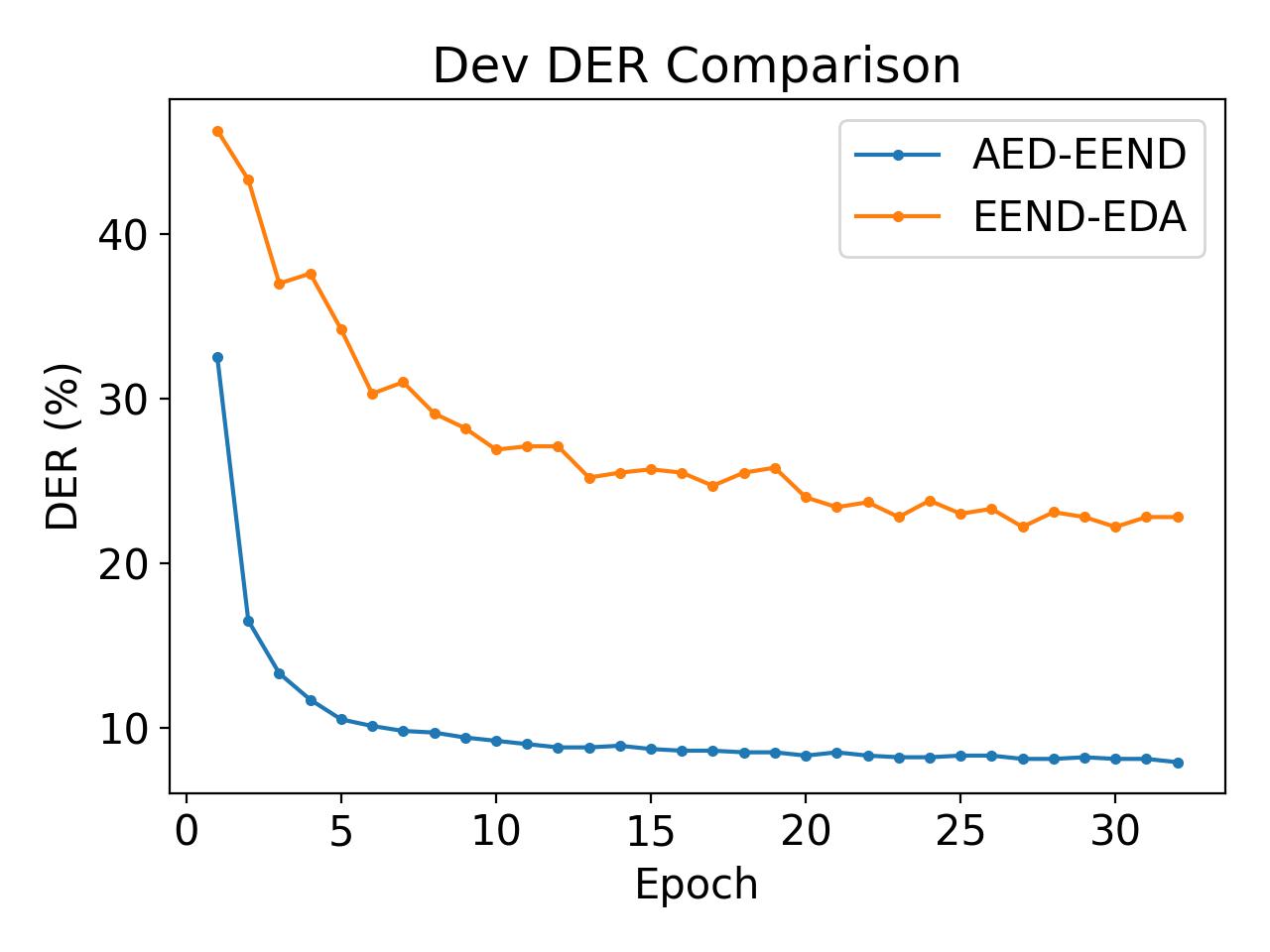}  
  \caption{DER (\%) comparison}
  \label{fig:der_compare}
\end{subfigure}

\caption{The convergence speed comparison between EEND-EDA and our proposed AED-EEND in the first 30 epoches. Both systems are trained on the SM dataset with 1-5 speakers.}
\label{fig:aed_eda_training_comparison}
\end{figure}

As introduced in section \ref{ssec:training_setup}, we directly trained our AED-EEND system on the variable number-of-speaker simulation dataset, whereas the EEND-EDA \cite{horiguchi2022encoder} system is first trained on the 2-speaker simulation data set and then further trained on the simulation set with more speakers. This additional step was necessary for the EEND-EDA system due to its utilization of permutation invariant (PIT) loss, which poses challenges during training. The pre-training on the 2-speaker simulation dataset enabled the EEND-EDA model to better adapt to datasets with a flexible number of speakers. In our case, our system does not encounter speaker permutation issues, which facilitates easier convergence during training. We validate this assumption in Figure \ref{fig:aed_eda_training_comparison}, where we compare the training progress of our AED-EEND system and the EEND-EDA system on the SM dataset with 1-5 speakers. The plotted loss and diarization error rate (DER) statistics for the development set clearly demonstrate that our AED-EEND system achieves significantly faster convergence compared to the EEND-EDA system. This observation implies that we can employ simpler training strategies to achieve superior diarization performance.

\begin{table*}[ht!]
\footnotesize
\centering
\caption{\textbf{\textsc{CallHome} DER (\%) results with optimized setup.}
}
\begin{adjustbox}{width=.8\textwidth,center}
\begin{threeparttable}
\begin{tabular}{ll|l| cccccc}
\toprule
\multirow{2}{*}{Method} & \multicolumn{2}{c}{Simulation Data} & \multicolumn{6}{c}{spk\#}\\

\cmidrule(r){2-9}
 & Spk \# & Datatype & 2 & 3 & 4 & 5 & 6 & all\\
 \hline
EEND-EDA \cite{horiguchi2020end} & $k \in \{1, . . . , 4\}$ & SM & 8.50 & 13.24 & 21.46 & 33.16 & 40.29 & 15.29 \\
SC-EEND \cite{fujita2020neural} & $k \in \{1, . . . , 4\}$ & SM & 9.57 & 14.00 & 21.14 & 31.07 & 37.06 & 15.75 \\
MTFAD \cite{horiguchi2020end} & $k \in \{1, . . . , 4\}$ & - & - & - & - & - & - & 14.31 \\
AED-EEND  & $k \in \{1, . . . , 4\}$ & SM & 6.18 & 11.51 & 18.44 & 30.79 & 39.90 & 13.25 \\
AED-EEND-EE  & $k \in \{1, . . . , 4\}$ & SM & 6.93 & 11.92 & 17.12 & 28.22 & 31.97 & 12.91 \\
\hline
EEND-EDA \cite{horiguchi2022encoder} & $k \in \{1, . . . , 5\}$ & SM & 8.09 & 12.20 & 15.32 & 27.36 & 29.21 & 12.88 \\
AED-EEND & $k \in \{1, . . . , 5\}$ & SM & 6.26 & 11.54 & 17.28 & 29.05 & 30.61 & 12.56 \\

AED-EEND-EE & $k \in \{1, . . . , 5\}$ & SM & 6.42 & 11.47 & 17.20 & 27.15 & 29.36 & 12.43 \\
AED-EEND-EE (small) & $k \in \{1, . . . , 5\}$ & SM & 6.07 & 11.94 & 17.78 & 28.52 & 24.58 & 12.44 \\
AED-EEND & $k \in \{1, . . . , 5\}$ & SC & 5.84 & 11.02 & 15.32 & 27.68 & 27.19 & 11.61 \\
AED-EEND + Conformer & $k \in \{1, . . . , 5\}$ & SC & \textbf{5.58} & 10.49 & 13.06 & 26.47 & 24.06 & 10.66 \\
AED-EEND-EE & $k \in \{1, . . . , 5\}$ & SC & 5.69 & 9.81 & 12.44 & 23.35 & 21.72 & \textbf{10.08} \\


AED-EEND-EE (Init-Decode) & $k \in \{1, . . . , 5\}$ & SC & 5.83 & \textbf{9.66} & 13.73 & 24.46 & 25.74 & 10.57 \\
AED-EEND-EE (Random-Decode) & $k \in \{1, . . . , 5\}$ & SC & 5.66 & 9.68 & 14.62 & 24.35 & 22.44 & 10.55 \\
AED-EEND-EE +  Conformer & $k \in \{1, . . . , 5\}$ & SC & 5.61 & 9.78 & 13.56 & 26.50 & \textbf{20.04} & 10.35 \\
\midrule

VBx \cite{landini2022bayesian}  $^\dagger$ & - & - & 9.44 &  13.89 & 16.05 & 13.87 & 24.73 &  13.28 \\
EEND-post \cite{horiguchi2021end} $^*$  & - & - &  9.87 & 13.11 & 16.52 & 28.65 & 27.83 & 14.06 \\
EEND-vector clust.  \cite{kinoshita21_interspeech} $^*$  & - & - & 7.94 & 11.93 & 16.38 & 21.21 & 23.10 & 12.49 \\
EEND-GLA-Large \cite{horiguchi2022online} $^*$  & - & - & 7.11 & 11.88 & 14.37 &  25.95 & 21.95 & 11.84 \\
EEND-vector clust. + WavLM  \cite{chen2022wavlm} $^*$  & - & - & 6.46 & 10.69 & \textbf{11.84} & \textbf{12.89} & 20.70 & 10.35 \\
EEND-OLA + SOAP \cite{wang2023told} $^*$  & - & - & 5.73 & 10.31 & 11.96 & 23.89 & 20.39 & 10.14 \\





\bottomrule
\end{tabular}
\begin{tablenotes}\footnotesize
\item $^\dagger$: Oracle speech segments are used
\item $^*$: Two-stage systems
\end{tablenotes}
\end{threeparttable}
\label{table:enhance_callhome_res}
\end{adjustbox}
\end{table*}
\begin{table}[ht!]
\footnotesize
\centering
\caption{\textbf{DER (\%) results on the \textsc{CallHome} dataset with flexible number of speakers.} Systems are pre-trained on 1-4
speakers SM dataset.
}
\begin{adjustbox}{width=.47\textwidth,center}
\begin{tabular}{lcccccc}
\toprule
\multirow{2}{*}{Method} & \multicolumn{6}{c}{spk\#}\\
\cmidrule(r){2-7}
 & 2 & 3 & 4 & 5 & 6 & all\\
 \hline
 x-vector clustering \cite{horiguchi2020end} & & & & &  \\
 \hspace{3pt} Estimated \#spk & 15.45 & 18.01 & 22.68 & 31.40 & 34.27 & 19.43 \\
 \hspace{3pt} Oracle \#spk & 8.93 & 19.01 & 24.48 & 32.14 & 34.95 & 18.98 \\

 EEND-EDA \cite{horiguchi2020end} & & & \\
 \hspace{3pt} Estimated \#spk & 8.50 & 13.24 & 21.46 & 33.16 & 40.29 & 15.29 \\
 \hspace{3pt} Oracle \#spk & 8.35 & 13.20 & 21.71 & 33.00 & 41.07 & 15.43 \\



 AED-EEND & & & \\
 \hspace{3pt} Estimated \#spk & \textbf{6.18} & \textbf{11.51} & 18.44 & 30.79 & 39.90 & 13.25 \\
 \hspace{3pt} Oracle \#spk & 6.35 & 11.54 & 19.08 & 29.58 & 34.59 & 13.16 \\


   AED-EEND-EE & & & \\
 \hspace{3pt} Estimated \#spk & 6.93 & 11.92 & \textbf{17.12} & \textbf{28.22} & 31.97 & \textbf{12.91} \\
 \hspace{3pt} Oracle \#spk & 6.97 & 12.04 & 17.41 & 28.23 & \textbf{31.78} & 13.02 \\

 

\bottomrule
\end{tabular}
\label{table:flexible_spk_num_callhome_res}
\end{adjustbox}
\end{table}

\subsubsection{Results on CALLHOME Dataset}

We proceeded to evaluate our systems using the CALLHOME real dataset, and the results are presented in Table \ref{table:flexible_spk_num_callhome_res}. Interestingly, both the EEND-EDA system and our proposed AED-EEND system exhibit worse performance compared to the traditional x-vector clustering system in the 6-speaker condition. In \cite{horiguchi2020end}, the authors suggest that this discrepancy arises from the pre-training stage, where the models only encounter data containing four or fewer speakers. Surprisingly, our proposed AED-EEND-EE system demonstrates a noteworthy capability to generalize to unseen speaker number scenarios, surpassing the performance of the x-vector clustering system in the 5-speaker and 6-speaker conditions. However, compared to the results in Table \ref{table:flexible_spk_num_simu_res}, the overall performance gain from the Enhancer module becomes smaller and limited. We believe that our proposed Enhancer module does have strong modeling capabilities, but due to the mismatch between simulation data and real data, the model cannot generalize. The relevant experiments in the next section will also verify this point.



\subsubsection{Enhance the Results on CALLHOME Dataset with Optimized Setup}
\label{sssec:enhance_callhome}
To further enhance the performance on the CALLHOME dataset, we made modifications to the experimental setup used in Table \ref{table:flexible_spk_num_callhome_res} and present the updated results in Table \ref{table:enhance_callhome_res}. Firstly, we followed the approach outlined in \cite{horiguchi2022encoder} by incorporating a 5-speaker simulation dataset into our simulation training process. This change resulted in notable improvements for our proposed AED-EEND system in the evaluation sets with multiple speakers. Moreover, the results from Table \ref{table:flexible_spk_num_callhome_res} demonstrated that our AED-EEND-EE system already obtains the ability to generalize to situations with an unknown number of speakers. Therefore, the performance improvement after adding the 5-speaker simulation data was insignificant.
Additionally, our smaller version, AED-EEND-EE (small), introduced in section \ref{ssec:model_configuration_intro}, which has a similar parameter count to the EEND-EDA system, also outperformed the EEND-EDA system. Moreover, AED-EEND-EE (small) exhibited comparable performance to the AED-EEND-EE system, indicating that the dimension of the transformer has minimal impact on system performance.

Statistics in Table \ref{table:simulation_dataset} and \ref{table:real_recoding_dataset} highlight the significant difference in overlap ratio between the SM simulation data and the real data. To mitigate the mismatch between the simulation dataset and the real dataset, we replaced the SM dataset with the SC dataset, as described in section \ref{sssec:simu_data_intro}. This substitution led to further improvements in the performance of the AED-EEND system. Additionally, we explored the replacement of the transformer encoder with the Conformer, which yielded additional enhancements. Surprisingly, our AED-EEND-EE system, trained on the SC data, even outperformed the two-stage systems listed at the bottom of Table \ref{table:enhance_callhome_res}, achieving a new state-of-the-art performance on the CALLHOME evaluation set. Furthermore, we observed that the Enhancer module provided more significant improvements when trained on the SC data compared to the SM data. This phenomenon is consistent with the results in Table \ref{table:flexible_spk_num_simu_res} and Table \ref{table:flexible_spk_num_callhome_res}, where the Enhancer module exhibited significant improvements on the simulation evaluation set but yielded only marginal improvements on the real dataset. While we acknowledge the exceptional modeling capabilities of the Enhancer module, the substantial mismatch between the SM dataset and the real dataset limits its generalizability to the real dataset. Besides, we provide the results for the AED-EEND-EE (SC) system using the decoding methods Init-Decode and Random-Decode, where no explicit constraint is added to ensure there is one speaker in the enrollment area. Supervisingly, based on these two decoding methods, we also achieved a pretty good performance on CALLHOME evaluation set.

\begin{table}[ht!]
\centering
\captionsetup{justification=centering} 

\caption{\textbf{Speaker counting confusion matrix evaluated on CALLHOME dataset.}
All the systems are pre-trained on the simulation dataset with 1-5 speakers.} 

\begin{subtable}{0.24\textwidth}
\caption{EEND-EDA \cite{horiguchi2022encoder} \\ (Accuracy=84.4 \%)}   
\centering
\adjustbox{width=\textwidth}{%
\begin{tabular}{cc|cccccc}
\toprule 
& & \multicolumn{6}{c}{ Ref. \#Speakers } \\
& & 1 & 2 & 3 & 4 & 5 & 6 \\
\hline
\multirow{6}{*}{\rotatebox{90}{Pred. \#Speakers}}
& 1 & \textbf{0} & 1 & 0 & 0 & 0 & 0 \\
& 2 & 0 & \textbf{142} & 7 & 1 & 0 & 0 \\
& 3 & 0 & 5 & \textbf{54} & 4 & 0 & 0 \\
& 4 & 0 & 0 & 13 & \textbf{14} & 4 & 1 \\
& 5 & 0 & 0 & 0 & 1 & \textbf{1} & 2 \\
& 6 & 0 & 0 & 0 & 0 & 0 & \textbf{0} \\
\bottomrule
\end{tabular}%
}
\end{subtable}
\hfill
\begin{subtable}{0.24\textwidth}
\caption{AED-EEND (SM) \\ (Accuracy=73.6 \%)}   
\centering
\adjustbox{width=\textwidth}{%
\begin{tabular}{cc|cccccc}
\toprule 
& & \multicolumn{6}{c}{ Ref. \#Speakers } \\
& & 1 & 2 & 3 & 4 & 5 & 6 \\
\hline
\multirow{6}{*}{\rotatebox{90}{Pred. \#Speakers}}
& 1 & \textbf{0} & 5 & 0 & 0 & 0 & 0 \\
& 2 & 0 & \textbf{139} & 31 & 4 & 0 & 0 \\
& 3 & 0 & 4 & \textbf{42} & 14 & 3 & 2 \\
& 4 & 0 & 0 & 1 & \textbf{2} & 2 & 0 \\
& 5 & 0 & 0 & 0 & 0 & \textbf{0} & 0 \\
& 6 & 0 & 0 & 0 & 0 & 0 & \textbf{1} \\
\bottomrule
\end{tabular}%
}
\end{subtable}

  \vspace{0.5cm}
  
\begin{subtable}{0.24\textwidth}
\caption{AED-EEND-EE (SM) \\ (Accuracy=75.6 \%)}   
\centering
\adjustbox{width=\textwidth}{%
\begin{tabular}{cc|cccccc}
\toprule 
& & \multicolumn{6}{c}{ Ref. \#Speakers } \\
& & 1 & 2 & 3 & 4 & 5 & 6 \\
\hline
\multirow{6}{*}{\rotatebox{90}{Pred. \#Speakers}}
& 1 & \textbf{0} & 5 & 0 & 0 & 0 & 0 \\
& 2 & 0 & \textbf{141} & 25 & 4 & 0 & 0 \\
& 3 & 0 & 2 & \textbf{48} & 16 & 2 & 2 \\
& 4 & 0 & 0 & 1 & \textbf{0 }& 2 & 0 \\
& 5 & 0 & 0 & 0 & 0 & \textbf{0} & 1 \\
& 7 & 0 & 0 & 0 & 0 & 1 & 0 \\
\bottomrule
\end{tabular}%
}
\end{subtable}  
\hfill
\begin{subtable}{0.24\textwidth}
\caption{AED-EEND-EE (SC) \\ (Accuracy=77.6 \%)}   
\centering
\adjustbox{width=\textwidth}{%
\begin{tabular}{cc|cccccc}
\toprule 
& & \multicolumn{6}{c}{ Ref. \#Speakers } \\
& & 1 & 2 & 3 & 4 & 5 & 6 \\
\hline
\multirow{6}{*}{\rotatebox{90}{Pred. \#Speakers}}
& 1 & \textbf{0} & 4 & 0 & 0 & 0 & 0 \\
& 2 & 0 & \textbf{134} & 20 & 2 & 0 & 0 \\
& 3 & 0 & 10 & \textbf{53} & 10 & 0 & 1 \\
& 4 & 0 & 0 & 1 & \textbf{6} & 3 & 1 \\
& 5 & 0 & 0 & 0 & 2 & \textbf{1} & 1 \\
& 7 & 0 & 0 & 0 & 0 & 1 & 0 \\
\bottomrule
\end{tabular}%
}
\end{subtable}

\label{table:confusion_matrix}
\end{table}
\subsubsection{Speaker Counting Evaluation on CALLHOME}
Next, we assess the accuracy of our system in predicting the number of speakers and present the speaker counting confusion matrix in Table \ref{table:confusion_matrix}. The results reveal that our systems exhibit lower performance in the speaker counting task compared to EEND-EDA, indicating that the performance of speaker counting and speaker diarization may vary across different system types. Interestingly, we observe a positive correlation between the performance of diarization and speaker counting in our AED-EEND system. Perhaps in the future, we can enhance our AED-EEND system with the speaker counting ability form EEND-EDA to achieve better performance.

\begin{table}[ht!]
\footnotesize
\centering
\caption{\textbf{Diarization results on DIHARD II.}
}
\begin{adjustbox}{width=.47\textwidth,center}
\begin{threeparttable}
\begin{tabular}{l|cc}
\toprule
Method  & DER (\%) & JER (\%) \\
\hline
TS-VAD \cite{medennikov2020target} &  39.80 & 41.79 \\
TS-VAD (Multi-Channel) \cite{medennikov2020target} &  37.57 & 40.51 \\
SA-EEND \cite{horiguchi2022encoder} &  32.14 & 54.32 \\
EEND-EDA \cite{horiguchi2022encoder} & 29.57 & 51.50 \\
EEND-EDA (Iterative inference+) \cite{horiguchi2022encoder} &  28.52 & 49.77 \\
EEND-GLA-Large \cite{horiguchi2022online} &  28.33 & 50.62 \\
DIHARD II Track 2 Winner \cite{landini2019but} &  27.11 & 49.07 \\
\hspace{3pt} + EEND Post-Processing \cite{horiguchi2021end} &  26.88 & 48.43 \\

\hline
AED-EEND (Pre-trained on SM) &  27.06 & 51.72 \\
AED-EEND-EE (Pre-trained on SM) &  25.34 & 47.15 \\
AED-EEND &  25.92 & 49.53 \\
AED-EEND + Conformer $^*$ &  27.11 & 49.48 \\
AED-EEND-EE &  \textbf{24.64}     & \textbf{47.02} \\
AED-EEND-EE $^*$ &  26.13     & 47.28 \\
AED-EEND-EE + Conformer $^*$ &  25.12  & 47.56 \\

\bottomrule
\end{tabular}
\begin{tablenotes}\footnotesize
\item $^*$: the downsampling rate is set to 10 because the convolution operation in Conformer is sensitive to the time resolution.
\end{tablenotes}
\end{threeparttable}
\label{table:dihard_res}
\end{adjustbox}
\end{table}
\subsubsection{Results on DIHARD II Dataset}
In this section, we evaluate our system on the DIHARD II dataset, which offers a more challenging evaluation with a larger number of speakers and diverse conversation scenarios. The results are presented in Table \ref{table:dihard_res}. In the table, both our AED-EEND and AED-EEND-EE systems pre-trained on SM data share the same training data setup as the EEND-EDA system, yet both of our systems outperform the EEND-EDA system on the DER metric. In \cite{horiguchi2022encoder}, the iterative inference+ strategy is proposed to handle the problem of the number of outputs of EEND-EDA being empirically limited by its training dataset. It is worth noting that our AED-EEND-EE (SM) system outperforms the EEND-EDA (Iterative inference+) system a lot on both DER and JER metrics, which further improves that our AED-EEND-EE system can generalize better to the unseen number-of-speaker scenario. Then, we replace the SM training data with SC training data, similar to the results introduced in \ref{sssec:enhance_callhome}, and we achieve further improvement. Besides, we also tried to replace the transformer encoder with a Conformer in our DIHARD experiment and we use a different downsampling rate introduced in section \ref{ssec:evaluation_setup}. Remarkably, even with the change in downsampling ratio, the system utilizing the Conformer encoder yields excellent results. In Table \ref{table:dihard_res}, we also include the results of other works reported on the DIHARD II dataset, showcasing that our best-performing system achieves state-of-the-art performance on the DIHARD II evaluation benchmark.

\subsubsection{Results on AMI Dataset}

Finally, we evaluate our system on the AMI dataset, and the results are presented in Table \ref{table:ami_res}. Despite the AMI dataset consisting of only 3 or 4 speakers, the average duration of the AMI dev and test sets is 33 minutes, posing a significant challenge for the model's ability to handle long speech recordings. Similar to the results obtained on the CALLHOME and DIHARD II datasets, the Enhancer module and the utilization of the SC data consistently improve the system's performance. Remarkably, the Conformer encoder proves to be highly beneficial for the AMI dataset, further enhancing the system's capabilities and achieving state-of-the-art performance on the AMI evaluation benchmark.

\begin{table}[ht!]
\footnotesize
\centering
\caption{\textbf{Diarization results on AMI dataset.}
}
\begin{adjustbox}{width=.49\textwidth,center}
\begin{threeparttable}
\begin{tabular}{lcccc}
\toprule
\multirow{2}{*}{Method} & \multicolumn{2}{c}{Dev}  & \multicolumn{2}{c}{Eval}\\
\cmidrule(r){2-3} \cmidrule(r){4-5} 
 & DER (\%) & JER (\%) & DER (\%) & JER (\%) \\
 \hline
x-vec AHC + VB + OVL \cite{horiguchi2021end,perera2020speaker}  & - & - & 28.15 & 41.00 \\
\hspace{3pt} + EEND Post-Processing \cite{horiguchi2021end} & - & - & 27.97 & 40.57 \\
SA-EEND \cite{horiguchi2022encoder} & 31.66 & 39.20 & 27.70 & 37.50 \\
RPNSD \cite{huang2022joint} $^\dagger$ & - & - & 25.08 & 32.12 \\
Transcribe-to-Diarize \cite{kanda2022transcribe} & 23.51 & - & 24.43 & - \\
Multi-Class Spec-Clustering \cite{9383602} $^*$ & - & - & 23.60 & - \\
CmpEm + Overlap Detector \cite{9413752} & - & - & 22.93 & - \\
EEND-EDA \cite{horiguchi2022encoder} & 21.93 & 25.86 & 21.56 & 29.99 \\
Multi-scale SD \cite{park22d_interspeech} $^*$ & 22.20 & - &  21.19 & - \\
NSD-MA-MSE \cite{he2023ansd} & 16.71 & - & 16.95 & - \\

\hline 
AED-EEND (Pre-trained on SM) & 20.74 & 24.75 & 19.86 & 28.48 \\
AED-EEND-EE (Pre-trained on SM) & 18.89 & 23.19 & 16.91 & 25.32 \\
AED-EEND  & 18.94 & 22.93 & 18.77 & 25.89 \\
AED-EEND-EE & 15.89 & 19.57 & 16.33 & 23.73 \\
AED-EEND + Conformer  & 13.86 & 17.58 & 13.19 & 19.01 \\
AED-EEND-EE + Conformer & \textbf{13.63} & \textbf{17.17} & \textbf{13.00} & \textbf{18.52} \\
\bottomrule
\end{tabular}
\begin{tablenotes}\footnotesize
\item $^*$: Oracle speech segments are used
\item $^\dagger$: Oracle speaker number is used
\end{tablenotes}
\end{threeparttable}
\label{table:ami_res}
\end{adjustbox}
\end{table}

\begin{table*}[ht!]
\vspace{-5pt}
\footnotesize
\centering
\caption{\textbf{Speech type detection results.} ``Ours'' denotes our AED-EEND-EE + Conformer system trained on the 1-5 speakers SC simulation dataset. $\text{FA}$ represents the false alarm rate (\%) or false positive rate (\%). $\text{MISS}$ represents the miss error rate (\%) or false negative rate (\%). $\text{F}_1$ denotes the $\text{F}_1$-score (\%). For comparison with other systems, we inverted our system's prediction for non-speech to obtain the prediction for speech. The Silero-Vad model is a voice activity detection model, which can only distinguish speech and non-speech. The Pyannote toolkit only provides the interface for speech and overlap area prediction, and we derive the prediction for single-speaker speech area from these two predictions.
}
\begin{adjustbox}{width=.80\textwidth,center}
\begin{tabular}{c|c|ccc|ccc|ccc}
\toprule
\multirow{2}{*}{Dataset} & \multirow{2}{*}{System}  & \multicolumn{3}{c}{Speech} & \multicolumn{3}{c}{Single-Speaker} & \multicolumn{3}{c}{Overlap}\\
\cline{3-11}
                    && $\text{FA}$ & $\text{MISS}$ & $\text{F}_1$ & $\text{FA}$ & $\text{MISS}$ & $\text{F}_1$ & $\text{FA}$ & $\text{MISS}$ & $\text{F}_1$\\
\hline

\multirow{4}{*}{CALLHOME}
& Silero-Vad \cite{Silero-VAD} & 24.40 & 8.24 & 94.29 & - & - & - & - & - & - \\
& Pyannote \cite{Bredin2020,Bredin2021} & \textbf{17.10} & 7.07 & 95.34 & 38.38 & 11.73 & 87.65 & \textbf{3.58} & 57.59 & 52.12 \\
& Ours & 25.56 & \textbf{3.15} & \textbf{96.92} & \textbf{27.00} & \textbf{8.67} & \textbf{91.08} & 3.98 & \textbf{33.77} & \textbf{70.13} \\

\midrule
\multirow{3}{*}{DIHARD II}
& Silero-Vad & 40.50 & 9.87 & 88.23 & - & - & - & - & - & - \\
& Pyannote & 21.72 & 6.04 & 93.23 & 28.93 & 8.77 & 88.93 & \textbf{1.99} & 63.46 & 44.37 \\
& Ours & \textbf{13.87} & \textbf{3.91} & \textbf{95.64} & \textbf{16.56} & \textbf{8.61} & \textbf{91.68} & 3.03 & \textbf{36.67} & \textbf{61.37} \\

\midrule
\multirow{3}{*}{AMI Dev}
& Silero-Vad & 15.92 & 8.09 & 93.65 & - & - & - & - & - & - \\
& Pyannote & 11.12 & 5.05 & 95.91 & 18.65 & 8.42 & 91.30 & 2.19 & 35.34 & 70.97 \\
& Ours & \textbf{9.99} & \textbf{2.24} & \textbf{97.52} & \textbf{12.42} & \textbf{5.85} & \textbf{94.08} & \textbf{2.17} & \textbf{21.84} & \textbf{79.96} \\

\midrule
\multirow{3}{*}{AMI Eval}
& Silero-Vad& 9.14 & 14.19 & 91.27 & - & - & - & - & - & - \\
& Pyannote  & \textbf{8.85} & 5.79 & 95.96 & 20.37 & 8.54 & 91.13 & \textbf{1.66} & 41.68 & 68.30 \\
& Ours & 10.61 & \textbf{1.98} & \textbf{97.73} & \textbf{13.61} & \textbf{5.00} & \textbf{94.43} & 1.99 & \textbf{20.48} & \textbf{81.76} \\

\bottomrule
\end{tabular}
\label{table:silence_spk_overlap}
\end{adjustbox}
\vspace{-5pt}
\end{table*}


\section{Evaluation on Speech Type Detection Task}
\label{sec:speech_activity_res}

In this section, we evaluate the performance of our AED-EEND systems as an independent speech type detection model and compare it with the Pyannote \cite{Bredin2020,Bredin2021} and Silero-Vad \cite{Silero-VAD} systems, as displayed in Table \ref{table:silence_spk_overlap}. For Pyannote and Silero-Vad, we utilize their pre-trained models\footnote{\url{https://huggingface.co/pyannote/segmentation}}\footnote{\url{https://github.com/snakers4/silero-vad/tags}} directly. Meanwhile, for our system, in cases where the predictions of the three types of speech activities conflict with each other, we independently assess each prediction.
It should be noted that the Pyannote system in \cite{Bredin2021} is trained on AMI and DIHARD III datasets, while the training data used for Silero-Vad remains unclear. It is observed that our system achieves the highest F1-Score among the three systems for all the speech types. Although the comparisons may not be entirely fair due to the different training data configurations, these results still demonstrate the excellent speech type detection capability of our newly proposed AED-EEND-EE system. This ability significantly contributes to the overall success on the diarization task.

\section{Conclusion}
This paper proposed an innovative paradigm for speaker diarization by employing a simple attention-based encoder-decoder network for the task. Within this paradigm, we proposed a teacher-forcing training strategy to simplify the training pipeline and speed up the training process. We also proposed an iterative decoding method to output the diarization results for each speaker sequentially. 
Moreover, we propose a novel Embedding Enhancer module designed to enhance our system's ability to adapt to scenarios involving an unseen number of speakers. Recognizing a significant disparity between commonly used simulation datasets and real-world data, we advocate for the adoption of a more realistic simulation dataset to elevate the system performance.
Additionally, we explore replacing the transformer encoder with the Conformer model, aiming to better capture local information nuances. All these methodologies can be combined to boost the system performance, and it achieves the state-of-the-art across the established diarization benchmarks. Beyond diarization task, we identify our system's potential as a competitive speech-type detection model.



\bibliographystyle{IEEEtran}

\bibliography{mybib}

 





\end{document}